\documentclass[fleqn]{2020SCGE-my}
\usepackage{amssymb}
\usepackage{amsmath}
\usepackage{hyperref}
\usepackage{multirow,diagbox}
\usepackage{color}
\usepackage{setspace}
\usepackage{ulem}
\usepackage{enumerate}
\usepackage{mathtools}

 \makeatletter
 \newcommand{\pback}[1]{{
   \let\@rrow=\leftarrowfill
   \mathchoice{\AIN@stemPullBack{#1}{\@rrow}}{\AIN@stemPullBack{#1}{\@rrow}}
     {\AIN@indxPullBack{#1}{\@rrow}}{\AIN@indxPullBack{#1}{\@rrow}}}
   \vphantom{#1}}
 \newcommand{\AIN@stemPullBack}[2]{
   \vtop{\mathsurround=0pt
   \ialign{##\crcr$\textstyle{#1}\strut$\crcr
     \noalign{\kern-0.4ex\nointerlineskip}{\tiny#2}\crcr}}}

 \newcommand{\AIN@indxPullBack}[2]{
   \vtop{\mathsurround=0pt
   \ialign{##\crcr\hfil$\scriptstyle{#1}$\hfil\crcr
     \noalign{\kern+0.4ex\nointerlineskip}{\tiny#2}\crcr}}}

\makeatother

\begin{document}



\title{Entropy of black holes with arbitrary shapes in loop quantum gravity}{Entropy of black holes with arbitrary shapes in loop quantum gravity}

\author[1,2]{Shupeng Song}{}
\author[2]{Haida Li}{}
\author[1,2]{Yongge Ma}{mayg@bnu.edu.cn}%
\author[3]{Cong Zhang}{}

\AuthorMark{Song S P}

\AuthorCitation{Song S P, Li H D, Ma Y G, Zhang C}

\address[1]{School of Physics and Technology, Xinjiang University, Urumqi 830046, China}
\address[2]{Department of Physics, Beijing Normal University, Beijing 100875, China}
\address[3]{Faculty of Physics, University of Warsaw, Pasteura 5, 02-093 Warsaw, Poland}


\abstract{The quasi-local notion of an isolated horizon is employed to study the entropy of black holes without any particular symmetry in loop quantum gravity. The idea of characterizing the shape of a horizon by a sequence of local areas is successfully applied in the scheme to calculate the entropy by the $SO(1,1)$ BF boundary theory matching loop quantum gravity in the bulk. The generating function for calculating the microscopical degrees of freedom of a given isolated horizon is obtained. Numerical computations of small black holes indicate a new entropy formula containing the quantum correction related to the partition of the horizon. Further evidence shows that, for a given horizon area, the entropy decreases as a black hole deviates from the spherically symmetric one, and the entropy formula is also well suitable for big black holes.}



\maketitle


\begin{multicols}{2}
\section{Introduction}\label{sec:int}
As a remarkable prediction of general relativity (GR), the existence of black holes is supported by a lot of observational evidence, including the recent observation made by Event Horizon Telescope~\cite{Akiyama:2019cqa}. Theoretically, the Bekenstein-Hawking formula of black hole (BH) entropy~\cite{PhysRevD.7.2333,Hawking:1974sw} brings together the three pillars of fundamental physics, namely GR, quantum mechanics and statistical mechanics.  Thus, on one hand, it is generally expected that the statistical mechanical origin of BH entropy should be accounted for by a quantum theory of gravity~\cite{Rovelli:1996dv}. On the other hand, the computation of BH entropy from basic principles is an important test for any candidate theory of quantum gravity. While the definition of the event horizon of a BH is global and hence not suitable for
describing local physics, the notion of 
\Authorfootnote

\noindent an isolated horizon~(IH) is quasilocally defined~\cite{Ashtekar:1998sp}. It turns out that the thermodynamical laws of BH can be generalized to those of IH \cite{Ashtekar:2000hw}. Various attempts have been made to account for the entropy of certain IH by the well-known theory of loop quantum gravity (LQG)~\cite{Ashtekar:2004eh,rovelli2005quantum,thiemann2007modern,han2007fundamental}. Effort has also been made to calculate the entanglement entropy of an arbitrary boundary by LQG~\cite{Bodendorfer:2014fua}. In the schemes appearing so far~\cite{Ashtekar:2004nd,Ashtekar:1997yu,Ashtekar:2000eq,Beetle:2010rd,Engle:2009vc,Engle:2010kt,Wang:2014oua,Wang:2014cga,Pranzetti:2014tla,FernandoBarbero:2009ai}, one calculates the dimension of horizon Hilbert space compatible with a given macroscopic horizon area. All the treatments appearing so far depend essentially on the symmetries of IH, such as the spherical symmetry~\cite{Engle:2010kt} or the axi-symmetry~\cite{Ashtekar:2004nd}. In this paper, we propose a new and reasonable scheme for the computation of entropy of an IH with arbitrary shape in LQG.

Alternative to the Chern-Simons theory description of the IH degrees of freedom with either $U(1)$~\cite{Ashtekar:1997yu,Ashtekar:2000eq,Beetle:2010rd} or $SU(2)$~\cite{Engle:2009vc,Engle:2010kt} gauge group, the $SO(1,1)$ BF theory description of the horizon degrees of freedom can be applied to all dimensional IH~\cite{Wang:2014oua,Wang:2014cga}. Moreover, in comparison with the $SU(2)$ Chern-Simons approach where the area of IH has to be fixed in order to obtain the desired symplectic structure~\cite{Engle:2009vc,Engle:2010kt}, the area of the horizon is not fixed but encoded in the dynamical $B$ field in the BF approach~\cite{Wang:2014oua}. Hence, in the BF approach, the covariant phase space of the system is enlarged so that all spacetime solutions with any IH as an inner boundary are included. Note that for a spherically symmetric IH, a $SU(2)$ BF theory can also be obtained to describe the horizon degrees of freedom~\cite{Pranzetti:2014tla}. Alternative to the standard area operator~\cite{Rovelli:1994ge,Ashtekar:1996eg} in LQG, a flux-area operator was also proposed~\cite{FernandoBarbero:2009ai} for calculating the IH entropy. In the Chern-Simons approaches, the value $a_k$ of the area of IH in the Chern-Simons theory at level $k$ could not coincide with the spectrum of the standard area operator in LQG unless one introduced an area interval $[a_k-\delta,a_k+\delta]$ to ensure their consistency~\cite{Ashtekar:2000eq}, while the spectrum of the flux-area operator is evenly spaced and hence coincides with $a_k$ exactly~\cite{FernandoBarbero:2009ai}.  Taking account of the above facts, we will take the $SO(1,1)$ BF theory approach and employ the flux-area operator to study the entropy of an arbitrary IH in LQG.

According to their symmetries, isolated horizons can be classified into three categories: Type I (with spherical symmetry), Type II (with axisymmetry) and Type III (without symmetry other than the ``equilibrium" of the intrinsically geometric structures). To consider the entropy of an IH, the first challenge is how to characterize classically a general IH other than Type I, since the area can not determine uniquely its intrinsic geometry. While a natural attempt to address this issue is to define the geometric multipoles of a general IH, which is certainly difficult and still far from reaching~\cite{Ashtekar:2004gp}, we will take a new viewpoint that the local areas of its ``small enough" patches can characterize the intrinsic classical geometry of a general IH. The terminology ``small enough" can be understood as being macroscopically indistinguishable from a point but still containing huge degrees of freedom microscopically. Our description of the intrinsic geometry of IHs is  suitable for the IHs whose scalar curvature is positive almost everywhere.

The paper is organized as follows. In Sec.~\ref{sec:rev}, we will briefly review the $SO(1,1)$ BF theory description of the IH degrees of freedom. In Sec.~\ref{sec:ent}, we will first introduce how to characterize a general IH by an ordered area number sequence. Then the $SO(1,1)$ BF will be used to calculate the entropy of an IH with arbitrary shape in the framework of LQG. The results will be summarized and discussed in Sec.~\ref{sec:sum}. Throughout the paper, we use Latin alphabet $a,b,c,\cdots$ for abstract indices of spacetime, and capital $I,J,K,\cdots$ for internal Lorentzian indices.

\section{BF Theory Description of the Horizon Degrees of Freedom}\label{sec:rev}

Let us first recall the definition of an IH in GR. A 3-dimensional null hypersurface $\Delta$ equipped with an equivalence class $[l]$ of null normals $l^a$ in a spacetime with metric $g_{ab}$ is said to be an IH if the following conditions hold~\cite{Ashtekar:2000hw}.
\begin{enumerate}[(i)]
\item The topology of $\Delta$ is $S^2\times\mathbb{R}$ and the equivalence class $[l]$ of the future-directed $l$ is chosen by $l\sim l'$ if and only if $l'^a=C\,l^a$ for a positive constant $C$;
\item The expansion $\theta_{(l)}$ of $l$ vanishes on $\Delta$ for any null normal $l$;
\item Equations of motion hold on $\Delta$ and the stress-energy tensor $T_{ab}$ of matter fields at $\Delta$ is such that $-{T^a}_b l^b$ is future directed and causal for any future directed null normal $l$;
\item $[\mathcal{L}_l,\mathcal{D}]V=0$, for all vector fields $V$ tangential to $\Delta$ and all $l\in[l]$, where $\mathcal{D}$ is the uniquely induced covariant derivative on $\Delta$ inherited from $\nabla$. The actions of $\mathcal{D}$ on a vector field $X^a$ tangent to $\Delta$ and on an 1-form $Y_a$ intrinsic to $\Delta$ are given
by $\mathcal{D}_aX^b\widehat{=}\nabla_{\pback{a}}\tilde{X^b}$ and $\mathcal{D}_aY_b\widehat{=}\pback{\nabla_{a}\tilde{Y}_b}$ respectively, where $\tilde{X}^b$ and $\tilde{Y}_b$ are arbitrary extensions of $X^a$ and $Y_a$ to the 4-dimensional spacetime, the arrow under a covariant index denotes the pullback of that index to $\Delta$, and $\widehat{=}$ means equal on $\Delta$.
\end{enumerate}
Because of conditions (ii) and (iii) as well as the Raychaudhuri equation at $\Delta$, $l$ is  expansion, shear and twist free. Then, there exists
an one-form $\tilde{\omega}_a$ intrinsic to $\Delta$ such that $\mathcal{D}_a l^b\widehat{=}\,\tilde{\omega}_a l^b$, which implies that the  induced ``metric'' $\tilde{q}_{ab}\widehat{=}\,g_{\pback{ab}}$ on $\Delta$ satisfies $\mathcal{L}_l\tilde{q}_{ab}\widehat{=}\,0$. Condition (iv) ensures that $\mathcal{L}_l\tilde{\omega}_{a}\widehat{=}\,0$. Thus, the geometry of IH $(\Delta, [l])$ is completely specified by $(\tilde{q}_{ab},\tilde{\omega}_a)$ \cite{Ashtekar:2001jb}.

Consider a 4-dimensional spacetime region $\mathcal{M}$ with an arbitrary IH $\Delta$ as an inner boundary. The initial data locate on a spatial slice $M$ with the inner boundary $H=M\bigcap\Delta$. To derive the symplectic structure of the system, one starts from the Palatini action of GR on $\mathcal{M}$~\cite{Palatini1919ER,Ashtekar:2000hw},
\begin{align}
  S=\!-\frac{1}{16\pi G}\int_\mathcal{M}\!\Sigma_{IJ}\!\wedge\! F(A)^{IJ}\!+\!\frac{1}{16\pi G}\int_{\tau_{\infty}}\!\!\Sigma_{IJ}\!\wedge\! A^{IJ}\!,\label{eq:action}
\end{align}
where $G$ is the gravitational constant, $\Sigma_{IJ}\equiv\frac{1}{2}\epsilon_{IJKL}e^K\wedge e^L$ with $e^I$ being the co-tetrad, $A^{IJ}$ is the $SO(3,1)$ connection $1$-form, and $F^{IJ}$ is the curvature of $A^{IJ}$. Note that the boundary term at the spatial infinity $\tau_\infty$ in Eq.~\eqref{eq:action} is required by a well-defined action principle. To describe the geometry near $\Delta$, it is convenient to employ the Newman-Penrose formalism~\cite{Newman:1961qr} with the null tetrad $(l,n,m,\bar{m})$ adapted to $\Delta$ and $H$, such that the real vectors $\ell$ and $n$ coincide with the outgoing and ingoing future-directed null vectors at $\Delta$ respectively.
Then the co-tetrad fields are chosen as \cite{Kaul:2010kg}
\begin{align}\label{tetrad}
\begin{split}
e^0=\sqrt{\frac{1}{2}}(\alpha n+\frac{1}{\alpha} l),\ &e^1=\sqrt{\frac{1}{2}}(\alpha n-\frac{1}{\alpha} l),\\
e^2=\sqrt{\frac{1}{2}}(m+\bar{m}),\quad &e^3=i\sqrt{\frac{1}{2}}(m-\bar{m}),
\end{split}
\end{align}
where $\alpha$ is an arbitrary function of the coordinates, and now $l,n,m,\bar{m}$ denote the corresponding null co-tetrad. 
Thus, one local $SO(1,1)$ degree of freedom is left for the co-tetrad $\{e^I, I=0,1,2,3\}$. Restricted on the horizon, the co-tetrad satisfies $\pback{e}^0\,\widehat{=}\,\pback{e}^1$. Substituting it into the definition of $\Sigma_{IJ}$ and using the compatible condition between connection and co-tetrad, one can obtain the following identities for the pull-back forms on $\Delta$ \cite{Wang:2014oua,Wang:2015vua}
\begin{align}
&{\pback{\Sigma}}_{0i}\widehat{=} -{\pback{\Sigma}_{1i}},\quad \pback{A}^{0i}\widehat{=} \pback{A}^{1i},\quad \forall i=2,3,\nonumber\\
&\pback{A}^{01}\widehat{=}d\beta+\pi m+\bar{\pi}\bar{m}, \nonumber
\end{align}
where $\beta\equiv\kappa v+\ln\alpha$ with $\ell=\frac{\partial}{\partial v}$ and $\kappa$ being the surface gravity of the IH, the spin coefficients $\pi$ and $\bar{\pi}$ are the components of $\ell^a\nabla_a n$ along $\bar{m}$ and $m$ respectively.   

By the covariant phase space method~\cite{Lee:1990nz,Ashtekar:1991}, the horizon integral of the symplectic current can be calculated as
\begin{align*}
  \frac{1}{8\pi G}\int_{\Delta}\delta_{\left[1\right.}\!{\Sigma}_{IJ}\!\!\wedge\delta_{\left.2\right]}A^{IJ}= \frac{1}{4\pi G}\int_{\Delta}\delta_{\left[1\right.}\!{\pback{\Sigma}}_{01}\!\!\wedge\delta_{\left.2\right]}\pback{A}^{01}.
\end{align*}
The property of the IH ensures that $d\,\pback{\Sigma}_{01}\widehat{=}0$. Hence one can define a 1-form $B$ locally on $\Delta$ such that $\frac{1}{8\pi G}\pback{\Sigma}_{01}=d B$.

By performing an $SO(1,1)$ boost of $\{e^0,e^1\}$, one can show that $A^{01}$ is an $SO(1,1)$ connection and $d B$ is in its adjoint representation. In terms of Ashtekar-Barbero variables, the full symplectic structure can be obtained as \cite{Wang:2014oua,Wang:2015vua}
\begin{equation}
  \Omega(\delta_1,\delta_2)=\frac{1}{8\pi G\gamma}\int_M\! 2\delta_{[1}\Sigma^i\wedge\delta_{2]}\mathcal{A}_i+\oint_{H}\!2\delta_{[1}B \wedge \delta_{2]}A, \label{symp}
\end{equation}
where the conjugate pair for the bulk consists of the two-form $\Sigma^i$ determined by the co-triad field ${e^i}_a$ on $M$  by $\Sigma^i=\frac{1}{2}{\epsilon^i}_{jk}e^j\wedge e^k$ and the Ashtekar-Barbero $SU(2)$ connection $\mathcal{A}_i$ with $\gamma$ being the Immirzi parameter~\cite{Immirzi:1996di}, and the $SO(1,1)$ connection $A\equiv A^{01}$ on the boundary $H$
satisfies $d A=0$. Note that the remaining gauge group on $H$ is $SO(1,1)$, which is the essential gauge freedom of the tetrad adapted to $H$ and the null hypersurface $\Delta$. Note also that $d B$ is proportional to the volume element of $H$ up to orientation. Hence the boundary symplectic structure coincides with that of $SO(1,1)$ BF theory. In the quantum theory, to adapt the structure of LQG in the bulk, the IH degrees of freedom can be described by the quantum BF theory with the intersection points between $H$ and spin networks as sources.

Consider the graph $\Gamma$ underlying a spin network state intersects $H$ by $n$ intersections $\mathcal{P}=\{p_i|i=1,\cdots,n\}$. For every $p_i$ we associate a small enough neighborhood $s_i$. The physical degrees of freedom of the sourced BF theory are encoded in the variables $f_i=\int_{s_i}dB$. Then the corresponding quantum Hilbert space of the boundary BF theory reads $\mathcal{H}^{\mathcal{P}}_H=L^2(\mathbb{R}^n)$, and the bulk kinematical Hilbert space $\mathcal{H}^{\mathcal{P}}_M$ can be spanned by the spin network states $|\mathcal{P};\{j_p,m_p\},\cdots\rangle$ where $j_p$ and $m_p$ are respectively the spin label and magnetic number of the edge $e_p$ with an end point $p\in\mathcal{P}$. The integral $\Sigma^1(H)=\int_H\Sigma^1$ can be promoted as an operator in $\mathcal{H}^{\mathcal{P}}_M$ as
\[\hat{\Sigma}^1(\!H\!)|\mathcal{P};\!\{j_p,\!m_p\},\cdots\!\rangle\!=\!8\pi\gamma \ell^2_{P} \big(\!\!\!\sum_{p\in \Gamma\cap H}\!\!m_p\big) |\mathcal{P};\!\{j_p,\!m_p\},\cdots\!\rangle,\]
where $\ell_p=\sqrt{G\hbar}$ is the Planck length, while the eigenvalue of the flux-area operator $\hat{a}^{flux}_H$ on the same eigenstates reads \cite{FernandoBarbero:2009ai}:
\[a^{flux}=8\pi\gamma \ell^2_{P}\sum_{p\in \Gamma\cap H} |m_p|.\]
The space of kinematical states on a fixed $\Gamma$, satisfying the boundary condition, can be written as
\[\mathcal{H}_{\Gamma}=\bigoplus_{\{j_p,m_p\}_{p\in \Gamma\cap H}}\!\!\!\!\!\mathcal{H}^{\mathcal{P}}_M(\{j_p,m_p\}) \bigotimes\mathcal{H}^{\mathcal{P}}_H(\{\gamma m_p\}),\]
where $\mathcal{H}^{\mathcal{P}}_H(\{\gamma m_p\})$ denotes the subspace corresponding to the spectrum $\{\gamma m_p\}$ in the spectral decomposition of $\mathcal{H}^{\mathcal{P}}_H$ with respect to the operators $\hat{f}_p$.
Thus, the quantum states on $H$ could be labelled by sequences $(v_1,v_2,\cdots,v_n)$, where $v_i=2m_i$ are non-zero integers. For a given horizon area $a_H$, the sequences should satisfy
\begin{align}
 \sum_{p\in \Gamma\cap H} |v_p|=a,\quad v_p \in \mathbb{N}, \label{eq:cons-1}
\end{align}
where $a=\frac{a_H}{4\pi \gamma \ell_{P}^2}$. Moreover, for a BH, the spherical topology of $H$ imposes an additional restriction on $\Sigma^1(H)$ such that
\begin{align}
\sum_{p\in \Gamma\cap H}m_p=0,\label{eq:cons-2}
\end{align}
which is called the projection constraint~\cite{FernandoBarbero:2009ai}. Therefore, for a given horizon area $a_H$ of a spherically symmetric BH, the dimension $\mathcal{N}$ of the horizon Hilbert space is the number of sequences $(v_1,v_2,\cdots,v_n)$ satisfying constraints \eqref{eq:cons-1} and \eqref{eq:cons-2}. Thus the entropy of spherically symmetric BH with area $a_H$ can be calculated as
\begin{align*}
    S=\ln\mathcal{N}=\frac{\ln3}{\pi\gamma}\frac{a_H}{4\ell^2_{P}}-\frac{1}{2}\ln \frac{a_H}{4\gamma\ell^2_P}+\mathcal{O}(1).
\end{align*}
It should be noted that, in contrast to the $SU(2)$ Chern-Simons theory description of the horizon degrees of freedom~\cite{Engle:2009vc},
the $SO(1,1)$ BF theory description does not require any symmetry on the IH. This advantage enables us to calculate the statistic entropy of IHs with arbitrary shapes by the BF theory approach.

\section{Entropy of Isolated Horizons with Arbitrary Shapes}\label{sec:ent}
To distinguish different IHs with the same area, one needs to take the symmetry or shape of an IH into consideration. Using the foliation given in Ref.~\cite{Ashtekar:2001jb}, an IH ($\Delta$, $[\ell]$, $\tilde{q}_{ab}$, $\tilde{\omega}_a$) can be foliated into 2-spheres with a unique geometric pair $(q_{ab},\omega_a)$, where $q_{ab}$ and $\omega_a$ are respectively the projections of $\tilde{q}_{ab}$ and $\tilde{\omega}_a$ of $\Delta$ to the 2-sphere $H$. Thus the symmetry of an IH is determined by $q_{ab}$ and $\omega_a$.
Now we restrict our discussion to the non-rotating IHs, so that, only the information of $q_{ab}$ needs to be considered to distinguish the horizons with different shapes. Moreover, we only consider the IHs whose scalar curvature is positive almost everywhere. In this case, given a 2-sphere with a 2-metric $q_{ab}$, it can be globally immersed in the 3-dimensional Euclidean space $\mathbb{R}^3$, and $q_{ab}$ can completely determine the extrinsic curvature (i.e., the extrinsic shape) of the 2-sphere immersed in $\mathbb{R}^3$~\cite{spivak1970comprehensive}. To characterize the information of the 2-metric $q_{ab}$, we notice that any Riemann metric on a 2-sphere $H$ is conformal to a round metric~\cite{de2016uniformization}.
Moreover, if the scalar curvature of the physical metric $q_{ab}$ is positive almost everywhere, there exists a unique fiducial round metric $\mathring{q}$ on $H$\footnote{Private communications with A. Ashtekar and N. Khera.}, which is conformal to $q_{ab}$, i.e., $q_{ab}=\Omega^2\mathring{q}_{ab}$, where the conformal factor $\Omega$ is a positive function. Thus the information of $q_{ab}$ and hence the shape as well as the distortion of $H$ are fully reflected by the function $\Omega$, which is proportional to the area element on $H$.
To regularize the area element, one can divide $H$ into ``small enough" patches $\{O^{(i)}\}$ such that each $O^{(i)}$ has the same area measured by $\mathring{q}$. For instance, the partition can be realized by triangulation. The total number $K$ of the patches should satisfy $1\ll K\ll \frac{a_H}{4\pi\gamma \ell_p^2}$, where $a_H$ is the area of $H$ measured by $q$. Let $4\pi\gamma \ell_p^2\, a^{(i)}$ be the physical area of $O^{(i)}$. By fixing once and for all a way to order the patches $\mathcal{O}\equiv\{O^{(1)},O^{(2)},\cdots,O^{(K)}\}$, we obtain a corresponding ordered area number sequence $\{a^{(1)},a^{(2)},\cdots,a^{(K)}\}$ with $\sum_{i=1}^K a^{(i)}=\frac{a_H}{4\pi\gamma \ell_p^2}$, which is called a ``shape" of $H$ with the total area $a_H$ and almost everywhere positive scalar curvature. It is obvious that the shape of an IH can really be reconstructed by the area number sequence by the following steps in our cases. First, one fixes an ordered partition for the IH with a given total area. Second, one assigns the area numbers to the patches according to their orders. Thus, different area number sequences would give different conformal factors and hence the  information of physical metric $q$ including the extrinsic shape or distortion of the horizon. The differences among the area numbers within one sequence would reflect the distortion of the IH.

\begin{figure}[H]
  \centering
  \includegraphics[width=8cm]{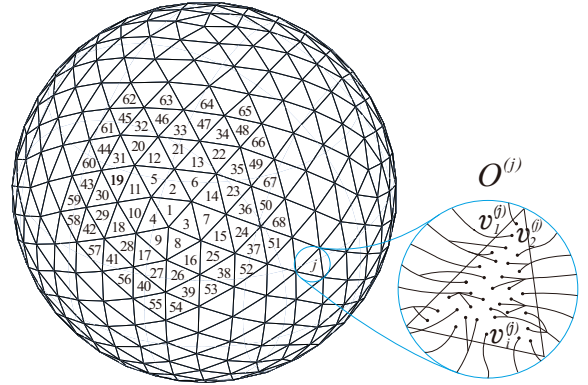}\\
  \caption{Ordering patches of $H$ and their intersections with spin networks}
\label{triang}
\end{figure}

A way to assign the ordering of the patches is shown in Fig.~\ref{triang}. The ordering is given as follows. We first choose an arbitrary patch and number it by 1, and number one of its neighbors by 2. Then we give the next number to the unnumbered neighbor of the previous patch with the smallest number clockwise, and repeat the last step until all patches are numbered. Once the ordering of the patches is fixed, if one exchanged the positions of two elements in the number sequence, the patches corresponding to these positions could have different areas. Even if two different orders of the same area number sequence are diffeomorphism equivalent on $H$, they still can be distinguished by external geometry, since the intersections attached the external geometry in different ways~\cite{Rovelli:1996dv}. Therefore, the different orders of the same area number sequence represent different shapes. The intersections inside a patch $O^{(j)}$ contribute the sequence $v^{(j)}\equiv(v^{(j)}_1,v^{(j)}_2,\cdots,v^{(j)}_i,\cdots)$ to $O^{(j)}$.

To calculate the entropy of an IH with a given shape, we will trace out the degrees of freedom corresponding to the bulk but take account of the horizon degrees of freedom~\cite{Ashtekar:2000eq}. It should be noted that one only needs to consider the diffeomorphism equivalence class of the ordered patches $\mathcal{O}$ as well as the intersections in each patch, while the possible positions of the patches and the intersections are irrelevant. As in the usual treatment in LQG~\cite{Ashtekar:2000eq}, we assume that for each given ordered sequence $(m_1,m_2,\cdots,m_n)$, there exists at least one state in the bulk Hilbert space $\mathcal{H}^{\mathcal{P}}_M$, which satisfies the Hamiltonian constraint. Then the dimension of the boundary Hilbert space $\mathcal{H}_H$ is given by the number of ordered sequences $(v^{(1)}_1,v^{(1)}_2,\cdots;v^{(2)}_1,v^{(2)}_2,\cdots;\cdots;v^{(K)}_1,v^{(K)}_2,\cdots)$ subject to the following piece-area constraints and projection constraint
\begin{eqnarray}
  \sum_i\left|v^{(j)}_i\right|&=&a^{(j)},\quad \forall j, \label{AreaCons}\\
  \sum_{j=1}^K\, V^{(j)}&=&0, \label{ProjCons}
\end{eqnarray}
where a quantum number $V^{(j)}\equiv\sum_iv^{(j)}_i$ is defined for each patch $O^{(j)}$. Note that Eqs.~\eqref{AreaCons} and \eqref{ProjCons} imply that $\frac{a_H}{4\pi\gamma \ell_p^2}$ has to be an even positive number. It should be noted that, the feature of Eqs.~\eqref{AreaCons} and \eqref{ProjCons} ensures that the number of their solutions would not change if one reordered the area numbers. Hence although the different orders of the same combination of piece area numbers represent different shapes, they have the same microstate numbers and hence the same entropy.

To solve the above number-theoretic and combinatorial problem by the generating function method~\cite{Sahlmann:2007jt,Agullo:2008yv,Agullo:2010zz,BarberoG.:2008ue}, we define the one-step function of each patch $O^{(j)}$ by
\[f(x_j,z)= \sum_{n_j=1}^{\infty}(z^{n_j}+z^{-n_j})x^{n_j}_j.\]
Here we use the powers of the variables $x_j$ and $z$ to represent respectively the area number and magnetic number contributed by one intersection in the patch $O^{(j)}$. Since the magnetic number could  be either positive or negative, there are following two cases.
The term $z^{n_j}x^{n_j}_j$ represents that an intersection contributes to both the magnetic number and area number by $n_j$, while the term $z^{-n_j}x^{n_j}_j$ contributes to the magnetic number by $-n_j$ and the area number by $n_j$. Thus, the one-step function is the summation of all situations of how the intersection contributes to the piece area number and magnetic number. The $n$th power $f^n$ of the one-step function gives the summation of all situations of how $n$ ordered intersections contribute to the piece area number and magnetic number. Summing over all possible intersections, one gets the generating function $G(x_{\!j},z)$ for $O^{(j)}$.
It could be expanded by the power series of variables $x_j$ and $z$, whose powers represent piece area number and magnetic number respectively. Thus, the generating function reads
\begin{align}
G(x_{\!j},z)&=1\!+\sum_{n=1}^{\infty}f^n(x_j,z)\nonumber\\
&=1\!+\!\!\!\sum_{a^{(j)}=1}^{\infty}\sum_{V^{(j)}=-a^{(j)}}^{a^{(j)}} \!\!\!\!\!\mathcal{N}(a^{(j)},V^{(j)})x^{a^{(j)}}_j\!z^{V^{(j)}},\nonumber
\end{align}
where the microstate number for given $a^{(j)}$ and $V^{(j)}$ is the coefficient
\begin{align}\nonumber
\mathcal{N}(a^{(j)},V^{(j)})=\begin{cases}
  A^n_i,& a^{(j)}=2n\!\!-\!\!1,\ V^{(j)}=\pm(2i\!\!-\!\!1); \\
  B^n_0,& a^{(j)}=2n,\quad\ V^{(j)}=0;\\
  B^n_i,& a^{(j)}=2n,\quad\ V^{(j)}=\pm2i;\\
 \,\,0, & \text{others.}
\end{cases}	
\end{align}
with $n,\ i\in \mathbb{N}^+$, $i\leqslant n$ and
\begin{subequations}\label{coeff}
\begin{align}
  A^n_i&=\sum _{j=0}^{n-i} \frac{(-3)^{j-1} (-6 n+2j+3) 2^{2 n-2 j-2} (2 n-j-2)!}{j! (n-i-j)! (n+i-j-1)!},\nonumber\\
  B^n_0&=\sum _{i=0}^{n} \frac{(-3)^{i-1} (-3 n+i) 2^{2 n-2 i} (2 n-i-1)!}{i! ((n-i)!)^2},\nonumber\\
  B^n_i&=\sum _{j=0}^{n-i} \frac{(-3)^{j-1} (-3 n+j) 2^{2 n-2 j} (2 n-j-1)!}{j! (n-i-j)! (n+i-j)!}.\nonumber
\end{align}
\end{subequations}
Since the microstate numbers of different patches are independent, the total generating function $G(\{x_{\!j}\},z)$ is the product of each $G(x_j,z)$ of $O^{(j)}$, i.e.,
\begin{equation}
G(\{x_{\!j}\},z)\!\!=\!\!\prod_{j=1}^K\!\frac{1}{1\!\!-\!\!f(x_{\!j},z)} \!\!=\!\!\prod_{j=1}^K\!\frac{(1\!-\!z x_{\!j})(z\!\!-\!\!x_{\!j})}{z\!\!-\!\!2x_{\!j}\!\!-\!\!2z^2x_{\!j}\!\!+\!\!3z x_{\!j}^2}.\label{gen}
\end{equation}
The total generating function \eqref{gen} could be expanded by the power series of the variables $z,x_1,\cdots,x_K$. The coefficient of $z^0\prod_{j=1}^K x_j^{a^{(j)}}$ in the expansion of Eq.~\eqref{gen} equals to the dimension of $\mathcal{H}_H$ satisfying \eqref{AreaCons} and \eqref{ProjCons} for the horizon $H$ with the given shape $\{a^{(j)}\}$. While the analytic calculation of this coefficient is difficult, we can employ the following operational scheme for its numerical computation.

For a given shape, one first compute the total microstate number for a possible quantum number sequence $\{V^{(j)}\}$ by neglecting the projection constraint as
\[\mathcal{N}\left(\{a^{(j)}\},\{V^{(j)}\}\right)=\prod_{j=1}^K \mathcal{N}(a^{(j)},V^{(j)}).\]
Then the total microstate number satisfying the projection constraint can be calculated by
\[\mathcal{N}\left(\{a^{(j)}\}\right)=\sum_{\{V^{(j)}\}}\left(P_{\{V^{(j)}\}}\cdot\mathcal{N}\big(\{a^{(j)}\},\{V^{(j)}\}\big)\right),\]
where the projection constraint is realized by demanding
\begin{equation}
P_{\{V^{(j)}\}}=
\begin{cases}
    0, & \sum^K_{j=1}V^{(j)}\neq0,\\
    1, & \sum^K_{j=1}V^{(j)}=0.
\end{cases}\nonumber
\end{equation}

\begin{figure}[H]
  \centering
  \includegraphics[width=8cm]{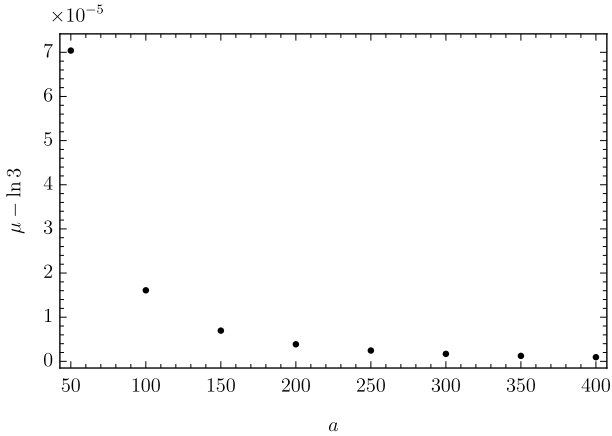}\\
  \caption{The difference between $\mu$ and $\ln 3$ with respect to $a$ for $K=5$ and $a\!\in\![50,400]$ with the interval 50.}
\label{alpha}
\end{figure}

To get an entropy formula by numerical calculation, we take the following ansatz in the light of the previous results in literatures~\cite{Wang:2014oua,FernandoBarbero:2009ai},
\begin{equation}\label{ansatz}
 S=\mu\, a+\sigma \ln a+\rho+\delta,
\end{equation}
where $a=\frac{a_H}{4\pi\gamma \ell_p^2}$ is the total area number, $\mu,\ \sigma$ and $\rho$ are undetermined constants, and $\delta$ represents the corrections related to different shapes. For the numerical simulation, our strategy is first to approximate the constants $\mu$ and $\sigma$ in the spherically symmetric case for a given $K$. Then we vary K and the shape to see whether the coefficients in Eq.~\eqref{ansatz} would be influenced. In the first step, for example, the difference between $\mu$ and $\ln 3$ is shown in Fig.~\ref{alpha} for $K=5$, and $a\!\in\![50,400]$ with the interval 50. It is obvious that the values of $\mu$ deviate from $\ln 3$ within the order of $10^{-5}$. Similarly, the numerical simulation approximates $\sigma$ to $-\frac{1}{2}$ within $10^{-3}$. In the second step, by varying $K$ and the shape, we find that $K$ influences the entropy at $\rho$ order, while the correction $\delta$ related to the shapes is at the order higher than $\rho$.

Thus, it turns out that the numerical computations of the entropy $S=\ln\mathcal{N}(\{a^{(j)}\})$ for small black holes (with area numbers less than 3200) indicate the formula:
\begin{equation}
  S_0=\frac{\ln 3}{\pi\gamma}\frac{a_H}{4\ell^2_P}-\frac{1}{2}\ln \frac{a_H}{4\gamma\ell^2_P}+K\ln\frac{2}{3}.\label{exp}
\end{equation}
This formula has the following convincing features. The coefficient $\frac{\ln 3}{\pi\gamma}$ of the leading order term in \eqref{exp} matches the results in other approaches employing also the flux-area operator~\cite{FernandoBarbero:2009ai,Wang:2014oua}. The coefficient $-\frac{1}{2}$ of the subleading logarithmic correction term matches the results of $U(1)$ Chern-Simons theory approaches~\cite{FernandoBarbero:2009ai,Domagala:2004jt}. The next order correction term containing $K$ matches the result of $SO(1,1)$ BF theory approach for spherically symmetric IH with $K=1$~\cite{Wang:2014oua}. Since the absolute error of entropy equals to the relative error of microstate number at first order approximation, i.e., \[\frac{\mathcal{N}-\mathcal{N}_0}{\mathcal{N}_0}=e^{S-S_0}-1\approx S-S_0\equiv\Delta S,\]
the validity of the formula~\eqref{exp} can be checked by its absolute error $\Delta S$ with the entropy $S$ of the numerical computation in various examples. For a spherically symmetric BH, every area number of the shape sequence takes the same value $a^{(j)}=\tilde{a}$. The numerical results of the entropy $S$ are compared with the error $\Delta S$ in Table~\ref{abserr} for different sizes of small black holes and different numbers $K$ of partition. The relative errors $\frac{\Delta S}{S}$ are in the order of $10^{-4}\sim10^{-5}$. It also indicates that, on one hand, for a given partition number $K$, the absolute error $\Delta S$ decreases in inverse proportion as the area number $\tilde{a}$ increases. On the other hand, for a given $\tilde{a}$, $\Delta S$ increases as $K$ increases. Further numerical computations indicate that there is an upper bound for $\Delta S$ as $K$ increases for a given $\tilde{a}$, and the upper bound of $\Delta S$ decrease as $\tilde{a}$ increases. Thus in the extreme case of $\tilde{a}=1$, the upper bound takes the maximal value $\Delta S_{max}\approx \frac{1}{2}\ln 2$.

\begin{table}[H]
\footnotesize
\caption{Numerical results of the entropy $S$ and error $\Delta S$  for spherically symmetric small black holes}
\label{abserr}
\begin{center}
\begin{tabular}{|c|c|c|c|c|c|c|}
  \hline
  $K$&  &$\tilde{a}=4$&$\tilde{a}=8$&$\tilde{a}=12$&$\tilde{a}=16$&$\tilde{a}=20$  \\ \hline
  \multirow{2}{*}{4}&$S$&14.0662&31.2578 &48.6221 &66.0511 &83.5144 \\
  \cline{2-7}
  &$\Delta S$&0.06890&0.02933 &0.01854 &0.01357 &0.01070 \\
  \hline
  \multirow{2}{*}{5}&$S$ &17.9502&39.5338 &61.2912 &83.1140 &104.9713\\
  \cline{2-7}
  &$\Delta S$&0.07553&0.03339 &0.02133 &0.01568  &0.01240 \\
  \hline
  \multirow{2}{*}{6}&$S$ &21.8526 &47.8287  &73.9798 &100.1965 &126.4481\\
  \cline{2-7}
  &$\Delta S$&0.08011&0.03607 &0.02317 &0.01708  &0.01352 \\
  \hline
  \multirow{2}{*}{8}&$S$ &29.6924&64.4551 &99.3940 &134.3991 &169.4392\\
  \cline{2-7}
  &$\Delta S$&0.08577&0.03940 &0.02547 &0.01882 &0.01493\\
  \hline
  \multirow{2}{*}{10}&$S$ &37.5621 &81.1123 &124.8396 &168.6332 &212.4621 \\
  \cline{2-7}
  &$\Delta S$&0.08914 &0.04138 &0.02684 &0.01987 &0.01577\\
  \hline
\end{tabular}
\end{center}
\end{table}
\begin{table*}[tb]
\caption{Numerical results of $S$ and $\Delta S$ for different sizes and shapes of small black holes}
\label{shapedif}
\begin{center}
\begin{tabular}{|c|c|c|c|c|c|}
  \hline
  Shape & $\Delta S /10^{-3}$ &Shape & $\Delta S /10^{-3}$  &Shape & $\Delta S /10^{-3}$\\ \hline
  (40,40,40,40,40) &6.064129780 &(80,80,80,80,80)&2.999874063&(160,160,160,160,160)&1.492091679\\
  (20,40,80,40,20) &6.064129778&(40,80,160,80,40)&2.999874063&(80,160,320,160,80)&1.492091679\\
  (15,35,100,35,15) &6.064129437&(30,70,200,70,30)&2.999874063&(60,140,400,140,60)&1.492091679\\
  (98,98,2,1,1)    &4.548119739&(196,196,4,2,2)&2.774008789&(392,392,8,4,4)&1.480203929\\
  (196,1,1,1,1)    &3.473723118&(396,1,1,1,1)& 1.727745157&(796,1,1,1,1)&0.861616306\\
  \hline
  \hline
  $S_0$ & 214.473608567 & $S_0$ &433.849492710 & $S_0$ & 872.947834587 \\
  \hline
\end{tabular}
\end{center}
\vspace{-0.5cm}
\end{table*}

The numerical results of $S$ and $\Delta S$ are compared for different sizes and shapes of small black holes in Table~\ref{shapedif} with the fixed partition number $K=5$. As expected, for a given total area, the spherically symmetric BH has the maximal entropy, and the entropy decreases as the BH deviates from the most symmetric one. It also indicates that the entropy difference between any two black holes with different shapes is within their absolute errors $\Delta S$. This explains why the entropy formula~\eqref{exp} does not depend on the shape of a BH, which should contribute higher order corrections than those in~\eqref{exp}. Further numerical computations indicate that the absolute error $\Delta S$ is at the order of $4\pi\gamma \ell^2_p K/a_H$ in the large area regime. Thus the entropy formula \eqref{exp} is also well suitable for big black holes. The numerical results of the entropy $S$ satisfy
\begin{equation*}
S_0-\ln\frac{3}{\sqrt{2\pi}}\leqslant S<S_0+\frac{1}{2}\ln 2.
\end{equation*}
The lower bound of $S$ occurs at $a_H=8\pi\gamma \ell_P^2$ and $K=1$, while the upper bound occurs at $K=\frac{a_H}{4\pi\gamma \ell_P^2}$ with $a_H\rightarrow\infty$.

\section{Discussion}\label{sec:sum}
Let us summarize with a few remarks. The key idea of this paper is to employ the ordered area number sequence $\{a^{(1)},a^{(2)},\cdots,a^{(K)}\}$ to characterize the shape of an IH. This characterization is valid for any horizon $H$ whose scalar curvature is positive almost everywhere. Since the entropy is only characterized by the area of the horizon in the classical thermodynamics of non-rotating IHs \cite{Ashtekar:2000hw} and type II IHs \cite{Ashtekar:2001is}, it is reasonable to assume that the entropy calculation method which we proposed can be applied to all types of IH. A delicate issue here is how to choose the partition number $K$ for a given $H$. Note that one of the motivations for partitioning $H$ as $\{O^{(i)}\}$ is that the area of a classical horizon cannot be generated by its one or several intersections with spin networks. Corresponding to a classically non-vanishing volume element of $H$, there must be at least one intersection with spin networks for each patch $O^{(j)}$ which is macroscopically indistinguishable from a point. Thus one reasonable choice is to ask the number $K$ to be proportional to the area $a_H$ and fix its value by assigning a mesoscopic scale $\delta$ to $\sqrt{\frac{a_H}{K}}$. Then the entropy formula \eqref{exp} becomes
\begin{equation}
S_0=\Big(\frac{\ln 3}{\pi\gamma}+\frac{4\ell^2_P}{\delta^2}\ln\frac{2}{3}\Big)\frac{a_H}{4\ell^2_P}-\frac{1}{2}\ln \frac{a_H}{4\gamma\ell^2_P}. \label{exp2}
\end{equation}
Since the introduction of $\delta=\sqrt{\frac{a_H}{K}}$ is due to the consideration of area contribution from quantum geometry, there is no reason to assume that the coefficient $\frac{4\ell^2_P}{\delta^2}\ln\frac{2}{3}$ should be included into the coefficient $\frac{1}{4}$ of the Bekenstein-Hawking entropy formula which concerns only classical geometry. Therefore, Eq.~\eqref{exp2} still suggests the Immirzi parameter as $\gamma=\frac{\ln 3}{\pi}$, while the very small number $\frac{4\ell^2_P}{\delta^2}\ln\frac{2}{3}$ can be regarded as a correction from quantum and semi-classical geometries. As a new quantum gravity effect, the latter might be fixed by other semi-classical consideration of LQG, for instance, the analysis in Ref.~\cite{Han:2016fgh}. It is interesting to note that the extreme choice of $K=\frac{a_H}{4\pi\gamma\ell_P^2}$ would give $\gamma=\frac{\ln 2}{\pi}$, which coincides with the lower bound of $\gamma$ obtained in Ref.~\cite{Domagala:2004jt}.

Although the shape of a BH was taken into account in our entropy calculation, it did not contribute to the entropy formula \eqref{exp} or \eqref{exp2} where the quantum corrections of logarithmic term and $K$ term were included. Our numerical computations indicate that, for a given total area, the entropy decreases as a BH deviates from the spherically symmetric one. Hence, the shape should contribute certain higher order correction to the entropy, which is worth further investigating.
It should be noted that, for type II IHs, the rotational 1-form $\omega_a$ which was ignored in our cases corresponds to the angular momentum \cite{Ashtekar:2004gp}, while the entropy corresponds to the area. In the first law of IH, the angular momentum does not affect the entropy. Hence it is very possible that $\omega_a$ is irrelevant to the entropy counting even at quantum level. To check this speculation, one still needs to realize the quantum degrees of freedom corresponding to $\omega_a$ for general IHs, which is an open issue deserving further investigation.

The entropy formula \eqref{exp} was speculated from the numerical calculation of small black holes and the consistency with the results in other approaches. In particular, the logarithmic correction term came from the imposition of the projection constraint \eqref{ProjCons}, and the $K$ term came from the partition of the horizon. One might be worried about that small black holes were employed for the numerical computation, since they could not be in equilibrium due to the Hawking radiation. However, our attitude here is to take the small black holes as ideal models to carry out the practical numerical computation. Both the analytic formula \eqref{gen} of the total generating function for calculating the BH entropy and the operational scheme for its numerical computation are well suitable for big black holes. Moreover, even the equilibrium could also be realized if one imagined a small BH inside an adiabatic box. Nevertheless, the analytic derivation of the entropy formula from the generating function \eqref{gen} is still an open issue which deserves studying further.

Though the $SO(1,1)$ BF theory description of the IH degrees of freedom was used in this paper, it is straightforward to apply our idea and scheme also to the Chern-Simons theory approaches. The $SO(1,1)$ BF theory approach with our new scheme can be extended to all dimensional IHs with the higher dimensional LQG in the bulk~\cite{Wang:2014cga,Bodendorfer:2011nv,Long:2019nkf}.

\Acknowledgements{We thank Abhay Ashtekar, Xiaokan Guo, Neev Khera, Jerzy Lewandowski, and Zhen Zhong for helpful discussions. YM and SS also thank Abhay Ashtekar for the generous hospitality and the China Scholarship Council for support during their visit to Penn State. This work is supported by NSFC with Grants No. 11875006 and No. 11961131013. CZ acknowledges the support by the Polish Narodowe Centrum Nauki, Grant No. 2018/30/Q/ST2/00811.}


\normalem
\providecommand{\href}[2]{#2}\begingroup\raggedright\endgroup

\end{multicols}
\end{document}